\documentstyle[12pt,aaspp4]{article}  

\begin{document}

\title{A Prescription for
Building the Milky Way's Halo \\ from Disrupted Satellites}

\author{Kathryn V. Johnston}
\affil{Institute for Advanced Study \\ Olden Lane, Princeton, NJ 08540 \\
email: kvj@sns.ias.edu}

\begin{abstract}

We develop a semi-analytic method for determining
the phase-space population of tidal debris along the orbit
of a disrupting satellite galaxy and 
illustrate its use with a number of applications.

We use this method to analyze Zhao's proposal that the
microlensing events towards the Large Magellanic
Cloud (LMC) might be explained by an appropriately placed
tidal streamer, and find that his scenarios lead either
to unacceptably high overdensities (10 -- 100\%) in faint star counts 
(apparent magnitudes 17.5 -- 20.5) away from the
Galactic plane or short timescales for the
debris to disperse ($10^8$ years).

We predict that the tidal
streamers from the LMC and
the Sagittarius dwarf galaxy currently
extend over more than $2\pi$ in azimuth along
their orbits.
Assuming that each satellite has lost half of its primordial mass,
we find that the streamers will have
overdensities in faint star counts 
of 10 -- 100\% and $<$1\% respectively, and conclude
that this mass loss rate is unlikely for the LMC, but
possible for Sagittarius.
If the Galaxy has accreted one hundred $10^5-10^6 M_{\odot}$ objects
(comparable to its current population of globular clusters)
at distances of 20 -- 100 kpc during its
lifetime then 10\% of the sky will now be covered
by tidal streamers.
\end{abstract}


\newpage
\section{Introduction}

Two fundamental astrophysical questions that can be addressed with
Galactic research are: how are galaxies formed?;
and what are they made from?
If galaxies formed hierarchically, we expect satellite accretion to
have played some role in Galactic history, and might hope to address the former
question by searching for signatures of these events.
The latter question is currently being addressed by microlensing
surveys of the Galactic halo (e.g. \cite{aea96}).
In most models for the currently available
microlensing results it is assumed that the halo
is sufficiently smooth for its global baryonic population to be 
determined from a local sample.
\cite{ms96} and \cite{z97} have pointed out that lumps in the halo,
either caused by dark clusters or debris from satellite disruption,
would jeopardize this interpretation.

To address both these questions, therefore, we need to quantify the
significance of the presence of
accreted structure in the Galactic halo.
\cite{uwg96} looked at the distribution 
of halo stars in age and metallicity and
concluded that no more than 
10\% of these stars could have been been accreted from the
destruction of galaxies with stellar populations like those of the
dwarf spheroidal satellites of the Milky Way. 
An alternative approach, 
which requires no assumptions about the age or metallicity of stars
in accreted satellites,
is to look for signatures of accretion
in the kinematic and spatial distribution
of the halo.

Numerical simulations have confirmed the physical intuition that
lumps observed in the halo's phase-space distribution 
could be associated with accretion events.
They have shown that an initially
spherical satellite will become distorted when introduced into 
a tidal field, with stripped stars forming
long streams along its orbit, both ahead and behind
(e.g. \cite{m90}; \cite{md94}; \cite{ola95}; \cite{pp95}; 
\cite{vw95}; \cite{jsh95}; Johnston, Hernquist \& Bolte 1996 --- hereafter
\cite{jhb96}; \cite{k97}).
Analyses of the tidal debris in these simulations have
demonstrated that these star streams can maintain 
coherence for the lifetime of the Galaxy, and
that long-lived moving groups of stars
are a natural consequence
of tidal disruption (\cite{jsh95}; 
\cite{jhb96}).

There is increasing evidence of morphological distortions, 
reminiscent of these simulations, in the Milky Way's satellites:
surface density contours of the Sagittarius dwarf galaxy
(discovered by \cite{igi94}) have now been reported to cover an
angular region $8\times 22$ square-degrees 
(\cite{m96}; \cite{a96}; \cite{fmrts96}; \cite{iwgis96});
distorted isophotes have been seen
in several globular clusters (\cite{g95});
and ``extra-tidal'' stars have been discovered
in most of the dwarf spheroidal satellites 
(\cite{ggrf92}; \cite{ih95}; \cite{ksh96}).

Substructure in the spatial distribution of the Milky Way's dwarf
spheroidal satellites, in the form of alignments along
Great Circles, has also historically been attributed to the ancient 
disruption of a much larger body
(\cite{lb76}; \cite{kd77}), and
further possible associations among the 
halo's globular cluster population have recently been found 
(\cite{lr92}; \cite{m94}; \cite{fp95}; \cite{lb295}).

In the stellar halo,
the advent of various digitized Plate Surveys
(e.g. \cite{apm}; \cite{poss1}; \cite{poss2}) raises 
the possibility of searching for 
lumps in its projected angular distribution:
the dwarf spheroidal galaxy Sextans was found as a simple overdensity
in star counts in the APM survey (\cite{i90});
\cite{kea97} have developed a circular filter technique that can be
used to search for systems with similar morphologies
and produce a quantitative
estimate of the completeness of the sample that has been found; 
and motivated by the work on satellite alignments \cite{jhb96}
proposed the method of Great
Circle Cell Counts to look for streams of stars along Great Circles
that might result from the disruption of a satellite galaxy.

Observers have been aware of moving groups as
distinct lumps in the local stellar phase-space distribution 
for a long time (e.g. \cite{e65})
and similar structures have also now been reported in
the halo (e.g. \cite{slc87}; 
\cite{ag92}; \cite{mmh94}, 1996; 
--- for a comprehensive review of this subject 
see \cite{mhm96}).
These discoveries are currently being 
followed up with systematic spatial and kinematic surveys 
of larger portions of the halo with the hope of assessing
how exceptional this substructure is (Majewski -- { private communication}).

Overall, the qualitative comparison of numerical and observational work 
outlined above clearly demonstrates that satellite accretion is 
ongoing in the Milky Way, and is likely to have occurred often
in the past.
To address the aim of quantifying the presence of accreted structure,
the characteristics of a specific phase-space feature (such as a
moving group)
or the frequency with which such features are found in
the stellar halo could be interpreted using a suite of
N-body simulations. 
However, the number of accretion events that such a study can survey is 
limited by time and computational cost considerations.
An alternative approach was adopted by Tremaine (1993), who
restricted the description of a trail to simple estimates for its
length and width given the parent satellite's
mass and distance from the Galactic center.
This allowed him to derive analytic expressions for the
fractional sky-coverage of tidal debris from an ensemble of satellites,
but at the expense of a more precise knowledge of the structure
of the individual trails.

In this paper, we develop a semi-analytic 
technique for finding the density distribution along 
the entire length of a debris trail, 
given a satellite of any mass and orbit, assuming only 
a disruption timescale.
This method is complimentary to both previous numerical 
and analytic work, being less computational expensive than the former
and hence able to cover a wider region of parameter space, while providing a
more detailed description of individual trails than the latter.
The success of the method is tested through comparison with N-body simulations.
As an example of its use
we discuss the implications of this model for
the surface density structure of debris trails in
the Milky Way system.
In a further paper we will extend the technique to generate 
full phase-space models of halos with
a variety of accretion histories,
and examine their observable properties (\cite{jm97}).

We present the method in \S 2,
apply it to the Milky Way system in \S 3,
and use it to test the plausibility of Zhao's (1997) model
of the microlensing results towards the LMC in \S 4.
We summarize and discuss the limitations of
this approach in \S 5.
The N-body simulations used to confirm the validity of our 
technique are outlined in Appendix A.

\section{Method}

In this section we present our method for constructing the phase-space
structure along a debris trail.
We outline our understanding of the debris' orbital energy distribution,
gained from studies of tidally disrupting hydrodynamic systems,
in \S 2.1.
In \S 2.2 we apply this picture to find the energy distribution 
of tidal debris from stellar systems, 
and then convert this to a density distribution along the 
tidal streamers by using the general properties of orbits in a halo potential.
This approach is tested by comparing its predictions with
the final particle positions in N-body simulations which have followed
satellites evolving in a Milky Way potential for 10 Gyrs (see Appendix A).
We summarize the technique in \S 2.3.

\subsection{Hydrodynamic Systems}

Analytic descriptions of tidal disruption and debris
dispersal have been developed for hydrodynamic systems
such as planetesimals encountering planets in the
Solar System (e.g. \cite{st92}) and stars falling into black holes 
(e.g. \cite{ek89}; \cite{k94}).
In these studies, the satellite is assumed to move on a parabolic orbit,
with internal properties such that it disrupts at pericenter.
The size of the satellite as it disrupts is the physical scale where
the internal and tidal forces balance --- roughly 
\begin{equation}
\label{rtideh}
	r_{\rm tide}=\left({m \over M}\right)^{1/3}R,
\end{equation}
where $m$ is the mass of the satellite system, $R$ is the distance
of closest approach between the parent and satellite
and $M$ is the mass of the parent system
({\it cf.} \cite{r47}).
The typical scale of the energy distribution of the debris is then the
change in the Keplerian potential energy on this spatial scale,
\begin{equation}
\label{epsh}
        \epsilon_{\rm Kepler} = r_{\rm tide}  {d\Phi \over dR} = 
\left( {m \over M} \right)^{1 \over 3} {GM \over R}.
\end{equation}
This energy scale lies between the orbital energy, 
$\epsilon_{\rm orb}=GM/R=(M/m)^{1/3}\epsilon_{\rm Kepler}$
and the internal energy of the satellite, 
$\epsilon_{\rm int}=Gm/r=(m/M)^{2/3}\epsilon_{\rm Kepler}$.
Thus, in the regime typically considered, where $(m/M)^{1/3}\ll 1$,
the debris from a disruption event occupies a region in energy
space where the perturbation from the satellite's
original orbital energy is small compared to $\epsilon_{\rm orb}$,
but large compared to $\epsilon_{\rm int}$.

Indeed, Sridhar \& Tremaine (1992) and Evans \& Kochanek (1989) both 
found the energy distribution of the debris to be 
continuous over a finite range, with scale $\epsilon_{\rm Kepler}$, 
centered on the initial orbital energy. 
Approximately half of the material moved to unbound, hyperbolic orbits
and half moved to bound elliptic orbits.
Ignoring the influences of self-gravity and fluid effects, 
the subsequent evolution of the debris 
could be  followed analytically
by simply mapping the distribution in energy to one in orbital
time-periods. 

\subsection{Stellar Systems}

The same physical principles used to determine
the energy distribution and dispersal of tidal debris
for hydrodynamic systems apply to stellar systems, but
the theory is less complex since it involves only collisionless
particle dynamics.
Its application to Galactic satellites differs in two ways:
(i) the parent potential is not Keplerian;
and (ii) the satellite undergoes repeated encounters.
Mass is lost continuously, but predominantly during the pericentric passages
(see e.g. \cite{jhb96}, their figure 2).
The implications of these differences are outlined in \S 2.2.1 and \S 2.2.2,
and the resulting
prediction for the matter distribution along a tidal streamer
is presented in \S2.2.3. 
The effect of the omission of self-gravity from
this analysis is discussed in \S2.2.4.

\subsubsection{Orbits in a Halo Potential}

Orbits in a spherical potential can be classified by their
energy $E$ and angular momentum $J$, which completely
specify their planar paths, $(R,\Psi)$ in
polar coordinates in the orbital plane, as well as their radial
and azimuthal time periods ($T_R$,$T_{\Psi}$).
Hence, for the purpose of characterizing the properties of orbits in the 
halo we consider 
a spherical approximation to the full Galactic potential that is used
in the numerical simulations (see Appendix A). 
This potential $\Phi_{\rm MW}$ is constructed from 
bulge and halo components taken directly from equation (\ref{mw})
and a third spherical component 
$\Phi=-GM_{\rm disk}/\sqrt{R^2+a^2}$, 
with the mass and scale length of the disk given in equation (\ref{mw}). 

For this spherical potential,
the extent of the energy distribution in the debris from a satellite
on orbit with pericenter $R_p$ can be found
analytically by using generalized
forms of equations (\ref{rtideh}) and (\ref{epsh})
\begin{equation}
\label{rtide}
	r_{\rm tide}=\left({m \over M_p}\right)^{1/3}R_p
\end{equation}	
\begin{equation}
\label{epsp}
	\epsilon_{\Phi}=r_{\rm tide}  {d\Phi \over dR} = 
\left( {m \over M_p} \right)^{1 \over 3} R_p  {d\Phi \over dR}
=  (GmR_p)^{1/3} \left({d\Phi\over dR}\right)^{2/3},
\end{equation} 
where $M_p=R_p^2(d\Phi/dR)/G$ is the mass of the Galaxy within $R_p$
\footnote{In a calculation analogous to Roche's (1847) work,
\cite{k62} found the limiting radius of a star cluster on
an orbit of eccentricity $e$ and pericenter $R_p$ to be
$r_{\rm tide}=R_p(m/M(3+e))^{1/3}$.}.
However, the azimuthal and radial orbital time periods must be found
numerically.

\begin{figure}
\begin{center}
\epsscale{0.5}
\plotone{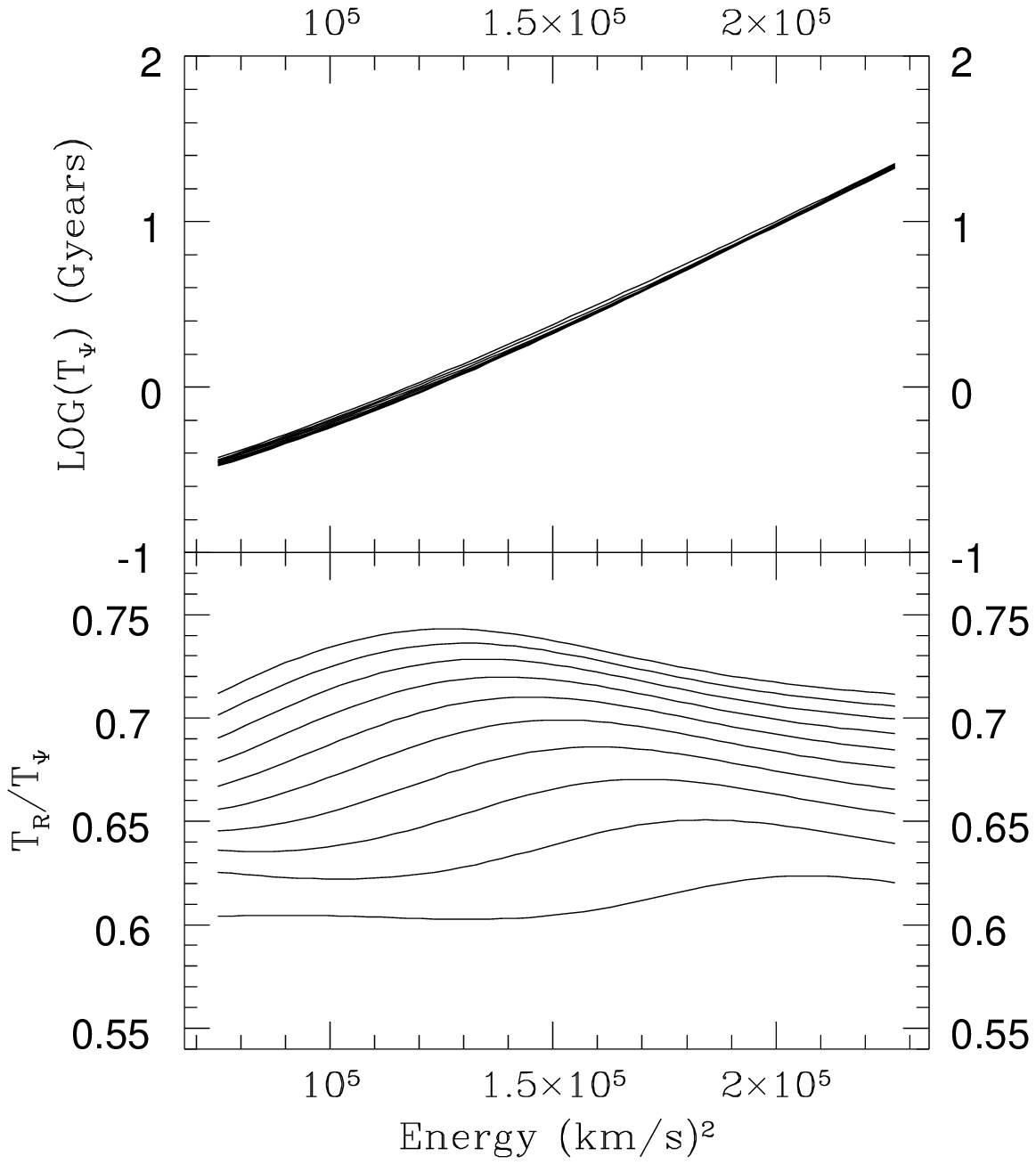}
\caption{The upper panel plots azimuthal time period 
$T_{\Psi}$ as a function of energy
for orbits with angular momentum $J/J_{\rm circ}=0.05-1.0$
in the spherical Milky Way potential $\Phi_{\rm MW}$
(eq. [\ref{mw}]).
The lower panel plots the ratio of radial to azimuthal time
periods $T_{R}/T_{\Psi}$ as a function of energy.
\label{tpsifig}
}
\end{center}
\end{figure}

The upper panel of Figure \ref{tpsifig} show $T_{\Psi}$
as a function of $E$ for $J/J_{\rm circ}=0.05, 0.1, 0.15, ... ,1.0$ (where
$J_{\rm circ}$ is the angular momentum of a circular orbit).
Note that the lines in this panel are almost indistinguishable
as $T_{\Psi}$ depends only very weakly on $J$.
For a circular orbit $T_{\Psi}$ can be calculated exactly, 
\begin{equation}
\label{tcirc}
	T_{\Psi}^{\rm circ}=2\pi \sqrt{R \over (d\Phi/dR)}.
\end{equation}
The solid lines in the lower panel of Figure \ref{tpsifig} show
the ratio $T_R/T_{\Psi}$ for the same values of $J/J_{\rm circ}$.
For a near-circular orbit,
$T_R$ was calculated using the epicyclic approximation
(see \cite{bt})
\begin{equation}
\label{tepi}
	T_{\rm R}^{\rm circ}={2\pi \over \kappa}, \qquad
	\kappa^2={d^2\Phi \over dR^2}+{3 J_{\rm circ} \over R^4}
		={d^2\Phi \over dR^2}+{3\over R}{d\Phi\over dR}.
\end{equation} 
For a radial orbit $T_R/T_{\Psi}=1/2$.

Since we are primarily interested in the construction of the outer
halo ($> 20$ kpc) from tidal debris, we might hope to represent
the full Milky Way potential as logarithmic in this
region, 
\begin{equation}
\label{log}
	\Phi_{\rm log}=v_{\rm circ}^2 \log(R/R_0), \,\,\, 
v_{\rm circ}=200 \> {\rm km/s}, \,\,\, 
R_0=\,\,{\rm arbitrary \,\, constant}.
\end{equation} 
(The I.A.U standard for the circular velocity at the solar radius
is $v_0=$220 km/s, but recent estimates tend to be lower than this ---
see \cite{penny} for a summary.) 
This further approximation is not strictly necessary, but it allows us
to express equations (\ref{epsp}), (\ref{tcirc}) and 
(\ref{tepi}) in simple form:
\begin{equation}
\label{eps}
	\epsilon_{\rm log}=\left({Gm \over R_p}\right)^{1 \over 3}
(v_{\rm circ}^2)^{2 \over 3}
\end{equation}
\begin{equation}
\label{tee}
        T_{\Psi}^{\rm circ}={2\pi \over v_{\rm circ}} R_{\rm circ}
={2\pi R_0 \over v_{\rm circ}} \exp \left( {E-v_{\rm circ}^2/2 \over v_{\rm circ}^2} \right)
=1.5\times \left ({R_{\rm circ} \over 50 {\rm \> kpc}}\right) {\rm
Gyrs},
\end{equation}
\begin{equation}
	{T_R^{\rm circ} \over T_{\Psi}^{\rm circ}}=\sqrt{1 \over 2}.
\end{equation}

\begin{figure}
\begin{center}
\epsscale{0.5}
\plotone{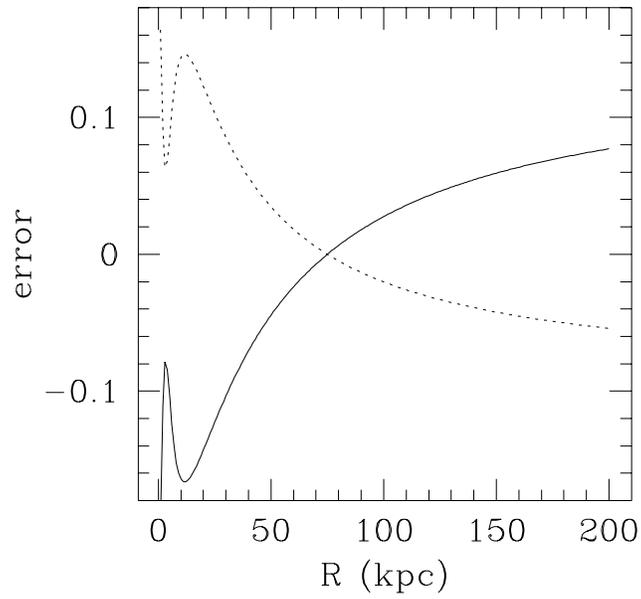}
\caption{
The fractional error in the calculation of $\epsilon$ (solid line -
see eq. [\ref{eps}]) and $T_{\Psi}$ (dotted line)
if our three-component Milky Way potential (see Appendix A)
is approximated as purely logarithmic at all radii.
\label{logfig}
}
\end{center}
\end{figure}

In Figure \ref{logfig}
we make some assessment of how successful this representation
is for our own three-component model of the Milky Way. The dotted 
line shows
$(\epsilon_{\rm log}/\epsilon_{\rm MW}-1)$ as a function of $R_p$,
and the solid line shows
$(T_{\rm log}/T_{\rm MW}-1)$ for time periods of 
circular orbits as a function of $R_{\rm circ}$,
where ``log'' and ``MW'' denote quantities calculated in
the potentials $\Phi_{\rm log}$ and $\Phi_{\rm MW}$ respectively.
The Figure demonstrates that these properties
in the two potentials differ by only 10\% in the region of interest.

We use Figures \ref{tpsifig} and \ref{logfig} to justify the following
approximations:
\begin{enumerate} 
	\item We use the logarithmic potential to
estimate the scale of the energy distribution (eq. [\ref{eps}]).
	\item We assume that the azimuthal time period $T_{\Psi}$
of an orbit is a function only of its energy,
independent of its angular momentum or eccentricity. 
We calculate $T_{\Psi}$
using the expression for a circular orbit of
energy $E$ in a logarithmic potential given
by equation (\ref{tee}).
This allows us to write down the relation between
$T_{\Psi}$ for orbits of energy $E$ and ($E+\Delta E$) as
\begin{equation}
\label{tau}
        T_{\Psi}(E+\Delta E)= \tau(\Delta E) T_{\Psi}(E), \, \, \,
	\tau(\Delta E)=\exp \left( {\Delta E \over v_{\rm circ}^2}\right).
\end{equation}
	\item We take $T_R/T_{\Psi}={\rm constant}=\sqrt{1/2}$.
Assuming this condition is strictly true, then the 
angular difference between successive
turning points (or pericenters and apocenters) is the same along all orbits.
Given the azimuthal time dependence $\Psi(t)$ of an orbit of energy $E$ 
and period $T_{\Psi}$, we can express
the azimuthal path of one
with energy ($E+\Delta E$) and period $\tau T_{\Psi}$ as
($\Psi_0+\Psi(t_0+t/\tau)$), where $\Psi_0$ and 
$t_0$ are constants chosen to take into account the
difference between the azimuthal and radial phases of the two
orbits at $t=0$.
\end{enumerate}

\subsubsection{Energy Distribution in Debris}

To simplify the calculation of the energy distribution in the tidal streamers
we model the disruption of a satellite as a series of discrete 
mass loss events occurring on each pericentric passage
(which will clearly be most appropriate for eccentric orbits).

\begin{figure}
\begin{center}
\epsscale{0.65}
\plotone{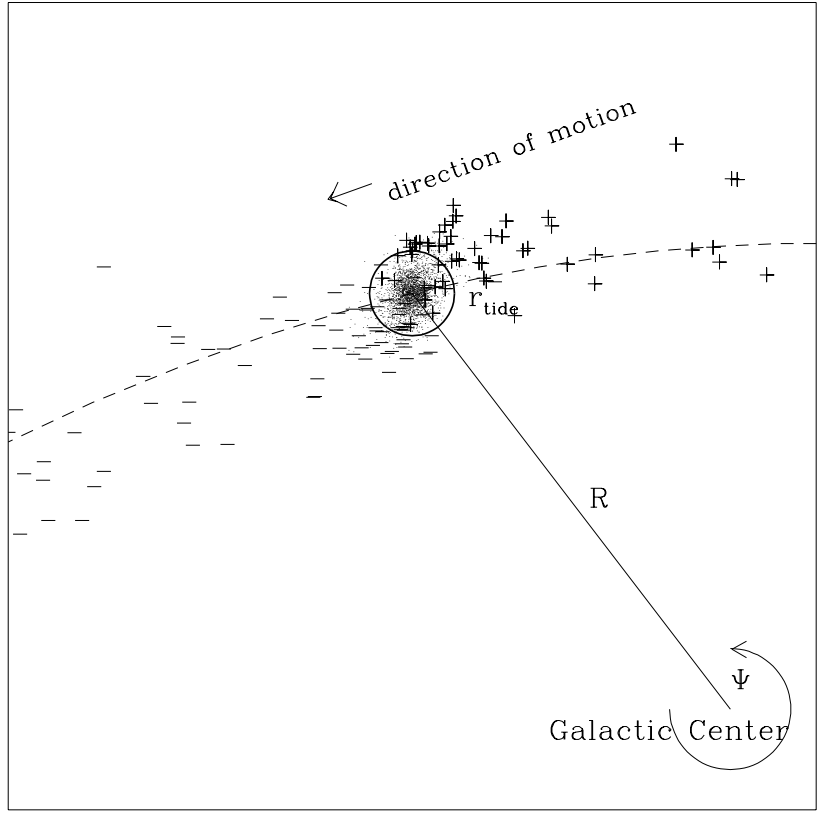}
\caption{
Snapshot of particle positions in the orbital plane of a simulation of a tidal
encounter. The dots label particles still bound to the satellite, and 
the ``-'' and ``+'' signs label those unbound on orbits with
lower and higher energy than the satellite orbit respectively. The dashed
line shows the satellite's path. The bold circle shows the physical scale
calculated from equation (\ref{rtide}).
\label{snapfig}
}
\end{center}
\end{figure}

Figure \ref{snapfig} is a ``snapshot'' from a simulated encounter,
showing the positions of
particles in the orbital plane of a satellite.
The dots in the plot show particles still bound to the satellite
and the bold circle shows the extent
$r_{\rm tide}$ from equation (\ref{rtide}) that sets the scale of 
the energy changes. 
The dashed line shows the satellite's orbit 
and the plus ($+$) and minus ($-$) 
signs show unbound particles which have moved
to orbits of higher and lower energies respectively.
The plot suggests that mass lost on passages prior to
disruption will make two distinct 
contributions to the energy distribution in the tidal streamers,
at positive and negative $\Delta E$, 
clearly separated from each other in energy by the region where
$\Delta E$ is too small for a particle to escape from the satellite.
In contrast, in their study of hydrodynamic
systems disrupting during a single encounter, Evans \& Kochanek (1989) 
found the
density of debris to be constant in energy across the energy scale
$\epsilon_{\rm Kepler}$ centered on the satellite's orbit
(i.e. $dN/d(\Delta E)=1/2\epsilon$ for $-\epsilon< \Delta E < \epsilon$).
In the light of this study, we expect the
particles which become unbound
on the final passage, as the satellite disrupts completely, to 
remain at energies close to the original orbit rather than
forming distinct populations at positive and negative $\Delta E$.

To convert these expectations to a formalism for the finding the
full energy distribution of mass lost
from a satellite that disrupts over several
orbits, we first define the energy scale 
of debris lost on the $i$th of $n$ pericentric passages to be
$\epsilon_i= (Gm_i/R_p)^{1/3}
(v_{\rm circ}^2)^{2/3}$, where $m_i$ is the mass
still bound to the satellite on that passage
({\it cf.} eq [\ref{eps}]).
Let $N_{\rm peri}(q_i)$ and $N_{\rm disrupt}(q_n)$ be
the fractional number densities in scaled energy 
$q_i=\Delta E/\epsilon_i$ for particles unbound on the $i$th
pericentric passage and as the satellite disrupts respectively. The full
distribution of tidal debris is given by
\begin{equation}
\label{ne}
        {dN \over d(\Delta E)}=\sum_{i=1,n} \left[
{\delta m_i \over m_{\rm debris}} 
{N_{\rm peri}(\Delta E /\epsilon_i)\over \epsilon_i} \right]
+{m_n \over m_{\rm debris}} {N_{\rm disrupt} (\Delta E/\epsilon_n)
\over \epsilon_n},
\end{equation}
where $m_{\rm debris}$ is the total mass of the debris and
$\delta m_i=(m_{i}-m_{i+1})$ is the mass lost on the $i$th passage.

\begin{table}
\begin{center}
\begin{tabular}{cccccc}
Model & $m$  		 & $R_p$ & $R_a/R_p$ & $n$	& $f$ \\
      & ($M_{\odot}$) & (kpc) 	 & 	 &      &	\\
\hline
\hline
1     & $1.0\times 10^9$ &	 39.6 & 3.4	&  5     & 0.18 \\
\hline
2     & $4.1\times 10^7$ & 	 43.4 & 4.0	 & 4	& 0.30 \\
\hline
3     & $1.9\times 10^7$ & 	 45.2 & 2.4	 & 1	& 0.65 \\ 
\hline
4     & $6.9\times 10^7$ & 	 26.4 & 3.0      & 5	& 0.30 \\
\hline
5     & $3.3\times 10^8$ &  	 43.4	& 4.0	 & 4	& 0.30 \\	
\hline
6     & $1.2\times 10^8$ & 	 43.4	& 4.0	 & 4	& 0.30 \\ 
\hline
7     & $1.4\times 10^7$ &  	 43.4	& 4.0	 & 4	& 0.30 \\
\hline
8     & $5.1\times 10^6$ &  	 43.4	& 4.0	  & 4	& 0.30 \\
\hline
9     & $5.8\times 10^7$ &  	 43.4	& 4.0	 & 2	& 0.50 \\
\hline
10    & $3.3\times 10^7$ & 	 43.4	& 4.0	 & 4	& 0.25 \\ 
\hline
11    & $2.9\times 10^7$ & 	 43.4	& 4.0	 & 4	& 0.20 \\ 
\hline
12    & $2.4\times 10^7$ &  	 43.4	& 4.0	 & 4	& 0.12 \\
\end{tabular}
\end{center}
\caption
{Parameters of models used to generate debris trails. Columns:
(1) model number; (2) satellite mass; (3) pericenter of orbit; (4) ratio
of apocentric to pericentric distance;
(5) number of pericentric passages either prior to destruction, or 
during simulation; (6) fraction of bound mass lost on each passage.
\label{modstab}
}
\end{table}

\begin{figure}
\begin{center}
\epsscale{0.75}
\plotone{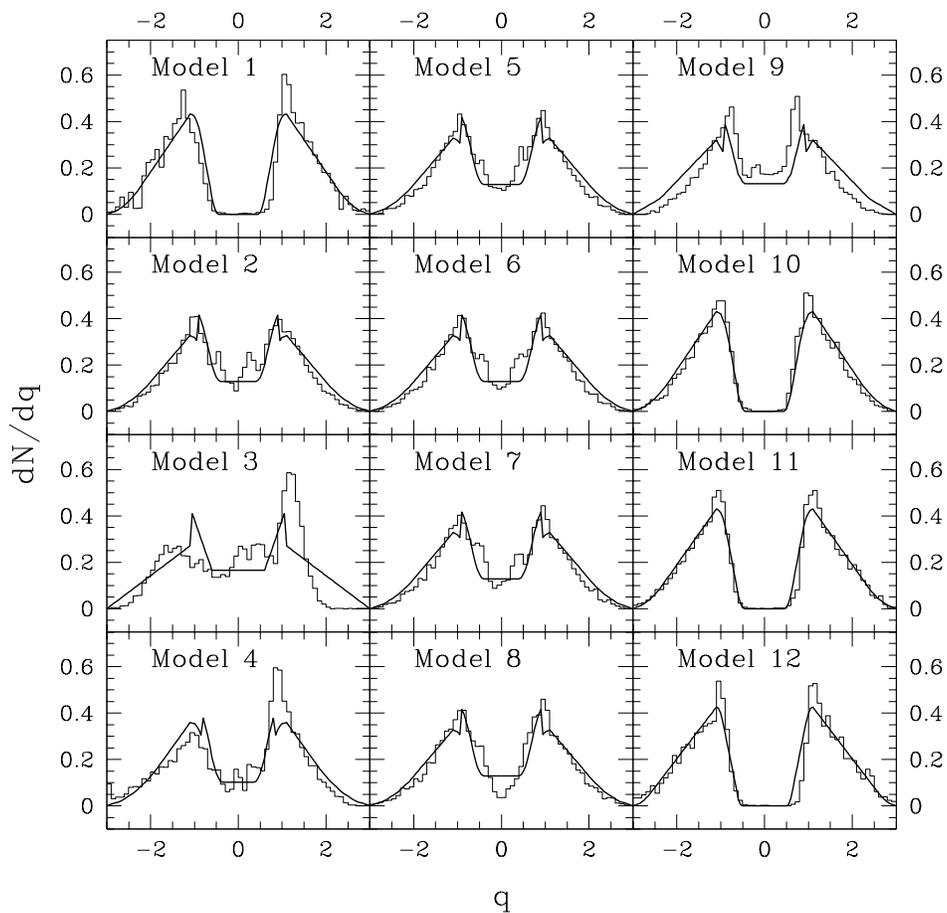}
\caption{Distribution of debris in scaled energy $q=\Delta E/\epsilon$
at the end of each simulation (histograms). The bold lines show the simple
analytic functions that we have used to approximate
each of these distributions (see eq. [\ref{ne}]).
\label{nefig}
}
\end{center}
\end{figure}

Figure \ref{nefig} shows the 
number density $dN/dq$ of unbound particles
in scaled energy changes $q=\Delta E/\epsilon$
at the end of each of the simulations (histograms - see Appendix A)
where $\epsilon$ is calculated from equation (\ref{eps}) with the values
for $m$ and $R_p$ listed in Table \ref{modstab}.
As predicted, the debris in all the models covers a similar range in
$q$ despite the variety of satellite orbits and masses 
employed.
In the models that were not disrupted during the course of the
simulation (Models 1, 10, 11 and 12), the debris shows two distinct
populations in energy changes, as expected from the discussion above.
The other models show similar distributions at lower and
higher energies formed at pericentric passages
prior to the destruction of the
satellite, along with a central population 
of particles with smaller $|\Delta E|$
from the final disruption event.
The histograms are similar enough 
to suggest that we can postulate forms for
intrinsic distributions in 
$q$ for the material torn during each pericentric passage as
\begin{equation}
\label{peri}
       {N_{\rm peri}(q)}= {1\over c-a} \times 
                \cases {(|q|-a)/(b-a), & if $a<|q|<b$; \cr
                      (c-|q|)/(c-b), & if $b<|q|<c$; \cr
                      0,             & otherwise, \cr}
\end{equation}
and from the satellite disruption as
\begin{equation}
\label{disrupt}
        {N_{\rm disrupt}(q)}= {1\over 2d} \times
		\cases {1, & if $|q|<d$; \cr
                      0, & otherwise, \cr}
\end{equation}
where $a=0.6, b=1.1, c=3.1$ and $d=1.5$.
The shape of these distributions and values for the parameters
are chosen by simple inspection of the histograms in Figure \ref{nefig}.

To verify the proposed forms for the scaled energy distributions
with each simulation
we need to specify the number of pericentric passages
$n$ and the mass of the satellite on each passage $m_i$,
and use these to convert $N_{\rm peri}$ and $N_{\rm disrupt}$
given by equations (\ref{peri}) and (\ref{disrupt})
to the full distribution in energy 
changes $dN/d(\Delta E)$ using equation (\ref{ne}). 
The mass loss history --- which we characterize by the values of $m_i$ ---
depends non-trivially on the mass, physical scale, and orbit of the
satellite,
and should be investigated with simulations if a detailed model of
a specific tidal disruption event is needed.
For our purposes of generating physically reasonable
debris trails, we assume that each satellite 
loses a constant fraction $f$ of its remaining mass instantaneously on each 
passage. To compare with the simulations
we choose $f$ and $n$ to approximate their mass loss histories 
by visual inspection of figure 2 in \cite{jhb96} (see Table \ref{modstab}).
Substituting $\delta m_i=f m_i$ and $m_i=(1-f)^i m$ in equation (\ref{ne})
we find the $dN/d(\Delta E)$
plotted as bold lines in Figure \ref{nefig}.
The fits are good enough to justify our use of the simple functional
forms for the energy distributions given by equations \ref{peri} and
\ref{disrupt}.

\subsubsection{Matter Distribution Along Tidal Streamers}

For any given  orbit, 
with azimuthal time dependence $\Psi(t)$,
we can generate a detailed description of the density
structure along the tidal streamers 
at any time by combining the energy distributions
and mass loss histories outlined in \S 2.2.2 with the approximations
for orbital properties given in \S 2.2.1.
We assume that particles lost at pericenter initially all have the same
radial phase as the satellite.
Then the problem of finding
the number density of particles at $\Psi$ from
mass lost with energy distribution
$dN/d(\Delta E)$ during the $ith$ pericentric passage at time $t_i$ 
reduces to a mapping (using eqs. [\ref{eps}], [\ref{tau}] and [\ref{ne}])
\begin{eqnarray}
\label{dndp}
        {dN_i \over d\Psi}&=&{dN_i \over d (\Delta E)}{d (\Delta E)\over d\tau}
{d\tau \over d\Psi(t_i+(t-t_i)/\tau)} \cr
                        &=&{v_{\rm circ}^2 \over \epsilon_i}{N(q_i)}
{1 \over u\dot{\Psi}(t_i+u)} \cr
			&=& \left({v_{\rm circ}^2 \over Gm_i/R_p}\right)^{1/3}
N(q_i){1 \over u\dot{\Psi}(t_i+u)},
\end{eqnarray}
where $u=(t-t_i)/\tau$, $q_i=\Delta E/\epsilon_i$ and
$N(q_i)$ is given by either equation (\ref{peri}) or (\ref{disrupt}),
as appropriate.
The complete distribution is calculated by summing over mass lost on each
of the pericentric passages.

\begin{figure}
\begin{center}
\epsscale{0.75}
\plotone{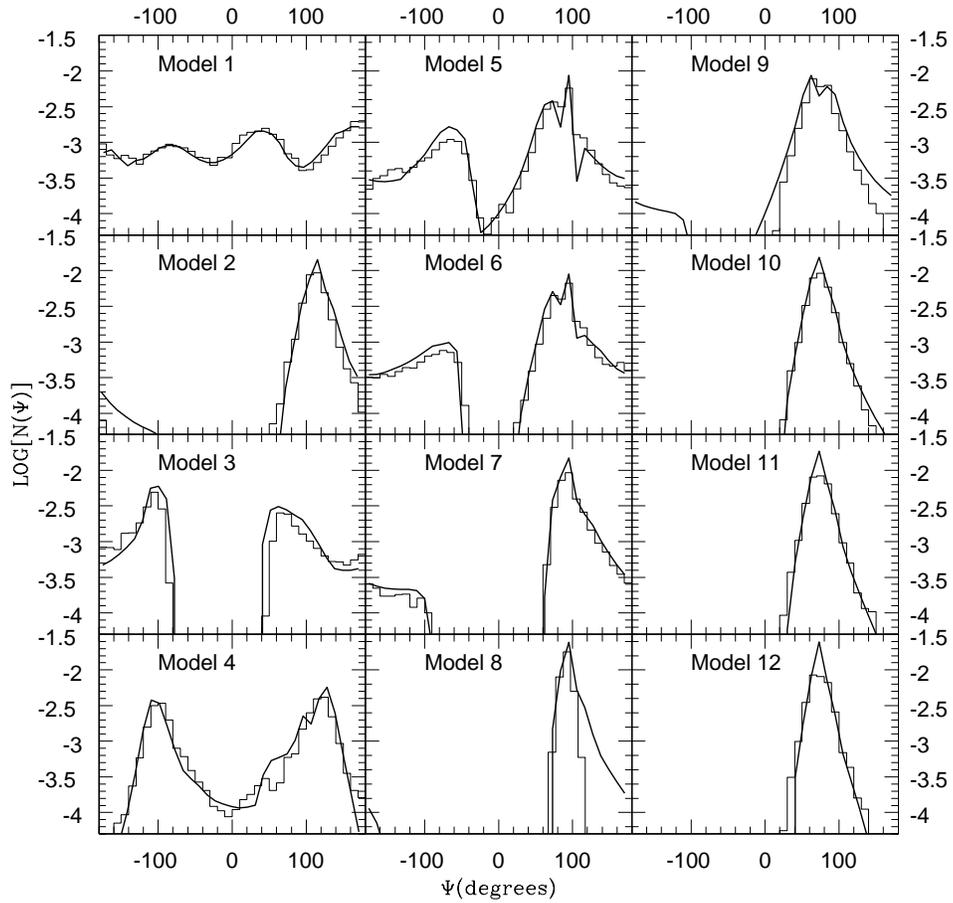}
\caption{Final distribution of debris along orbits for each simulation
(histograms). The bold lines are the predictions from equation
(\ref{dndp}).
\label{npsifig}}
\end{center}
\end{figure}

Figure \ref{npsifig} 
shows the fraction of unbound particles per degree, $dN/d\Psi$,
along the orbit of the satellite at the
end of each of the simulations (histograms).
The bold lines in the plot 
are the predictions of the mapping given in equation (\ref{dndp}), 
using the parameters
listed in Table \ref{modstab} and
taking $\dot{\Psi}$ to be the 
angular velocity along the (non-planar) path of the satellite
found by integrating its orbit
in the full Milky Way potential given in Appendix A.
In general, the semi-analytic approach reproduces the density
distribution to within a factor of two and over two orders of
magnitude in amplitude.

\subsubsection{Self-Gravity in Debris}

The gravitational interactions of particles in the debris trails are 
neglected in both the simulations (since the code uses spherical harmonic
expansions to evaluate the potential and cannot represent
the geometry of the streamers) and the above semi-analytic description.
However, we can argue that self-gravity is less important than
tidal forces for the predicted angular distribution by
comparing the influence of the two at each point along the streamers.

Following \cite{r47} and \cite{k62}, we expect
the internal and external forces to balance when the debris has a 
density of order
\begin{equation}
\label{crit}
	\rho_{\rm crit}={M(R) \over R^3}.
\end{equation}
The debris density $\rho$
can be calculated from equation (\ref{dndp}) by including
estimates for the stream's angular width $w$ and radial extent $\Delta R$.
The former is fixed by the range in orbital inclinations, set at 
pericenter to
\begin{equation}
\label{width}
	w= (r_{\rm tide}/R_p),
\end{equation}
and the latter is determined by the
scale of the energy changes, $\Delta R=\epsilon/(d\Phi/dR)=Rw$ ({\it cf.}
eq. [\ref{epsh}]).
Hence
\begin{equation}
\label{rho}
	\rho(R,\Psi)={m_{\rm debris}\over R^2 w \Delta R} {dN \over d\Psi}
={m_{\rm debris}\over R^3} \left({R_p \over r_{\rm tide}} \right)^2
{dN \over d\Psi}.
\end{equation}
For the center of a disrupting system, the
ratio of this density to the critical density for
self-gravity to be important is
\begin{equation}
	{\rho \over \rho_{\rm crit}}= {R_p \over R} {1 \over 2 d} 
{1\over t \dot{\Psi}}
\end{equation}
(combining eqs. [\ref{disrupt}], [\ref{dndp}], [\ref{crit}]) and
[\ref{rho}]).

\begin{figure}
\begin{center}
\epsscale{0.5}
\plotone{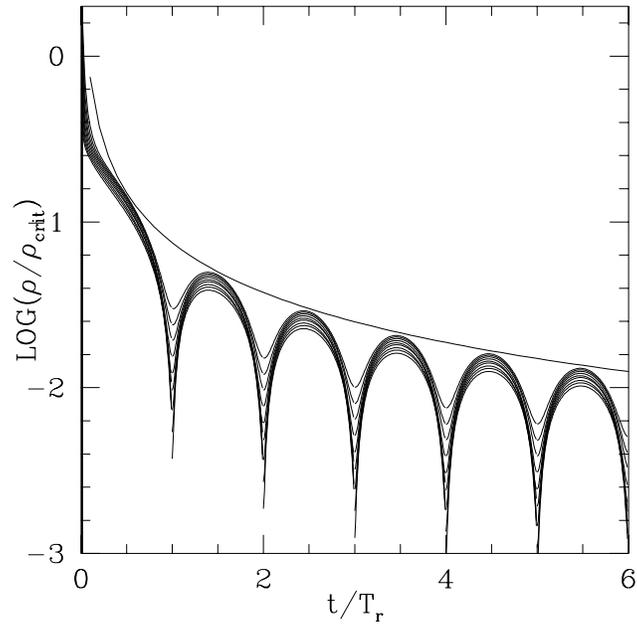}
\caption{Ratio of the density of a remnant
following the complete destruction of a satellite to
the critical density for self-gravity to be important.
The different curves are for orbits with the same energy
and $J/J_{\rm circ}=
0.28, 0.36...1.0$ or $R_a/R_p \sim 0.02-1.0$.
Pericenter corresponds  to the dip in each curve.
\label{densfig}
}
\end{center}
\end{figure}

This ratio is plotted in Figure \ref{densfig} for orbits with 
$J/J_{\rm circ}=0.28, 0.36 ... 1.0$ (or eccentricities
$R_p/R_a \sim 0.07 ... 1.0$, where $R_a$ is the apocentric distance).
At pericenter the curves reach minima
whose amplitudes decrease with $J$ as $\dot{\Psi}(R_p)$
increases.
At apocenter there is a corresponding decrease in $\dot{\Psi}(R_a)$
with $J$, but this is offset by an increase in the cross
sectional area of the trail because at large distances
it is less confined by the tidal field.
Thus the density evolution along a circular orbit (shown in bold)
serves as an upper limit
to the behavior along any orbit.
Since this  never exceeds the critical density we conclude 
that our prediction for the matter distribution along the streamers is
consistent with the neglect of
self-gravity in the analysis.
(The density becomes singular as $t\rightarrow 0$ since 
the initial angular extent of the debris in our description
is effectively zero at
this time.)

Note that mass escaping over the tidal radius
of a still-bound satellite will also have density
$\lesssim \rho_{crit}$, and is spread over a similar energy scale
({\it cf.} eqns [\ref{disrupt}] and [\ref{peri}]).
Hence our neglect of self-gravity 
is equally valid when modeling the behavior of
tidal streamers from Milky Way
satellites (such as the LMC).

\subsection{Summary of Technique}

Our method for generating debris distributions along tidal streamers,
can be summarized as follows:
\begin{description}
	\item{(i)} We assume that self-gravity is negligible
in the tidal streamers.
	\item{(ii)} We apply a simple physical picture to understand the 
energy scale, $\epsilon$, around the satellite's orbital energy
over which the debris is distributed 
(see \S 2.1, \S 2.2.1 and eq. [\ref{eps}]) 
and take a functional form for the full distribution
in scaled energy changes, $N(\Delta E/\epsilon)$, by examining the results 
from N-body simulations (see \S 2.2.2 and eqs. [\ref{peri}] and [\ref{disrupt}])
	\item{(iii)} We make the approximation that orbital time periods 
depend only on their energies.
Using this approximation, along with
the characteristics of orbits in a purely 
logarithmic potential (eq. [\ref{log}]), we relate the energy change $\Delta E$
to a change in orbital time period
(\S 2.2.1 and eq. [\ref{tau}]).
	\item{(iv)} We model mass loss as occurring instantaneously at the
satellite's pericentric passages and assume that the subsequent
turning points along the debris' and satellite's orbit 
are aligned in azimuth
(or $T_R/T_{\Psi}=$constant).
	\item{(v)} Combining the above points with the satellite's
orbit ($R(t),\Psi(t)$), we predict the debris distribution along the
path of the satellite from mass
lost at each previous pericentric passage, and sum these to find the
full density profile
(\S 2.2.3 and \S 2.2.4).
\end{description}

\section{Applications to the Milky Way System}

We illustrate some uses of the technique developed in the previous
section by asking: 
is it possible to detect tidal
streamers from any of the Galactic satellites? (\S3.1); 
how long will overdensities in star counts be noticeable
following a disruption event? (\S 3.2);
what is the angular scale of a tidal streamer and what
fraction of the sky might an ensemble of such streamers cover? (\S3.3).

\subsection{Should We Expect to See Tidal Streamers from Galactic Satellites?}

If we know the position, radial velocity and proper motion of a satellite,
we can integrate its orbit backwards in time in our (or
any other) Milky Way potential to find its angular path
$\Psi(t)$, pericentric position $R_p$
and number of encounters $n$ in the last 10 Gyrs.
Assuming the satellite, currently mass $m_{\rm sat}$, has lost mass 
$m_{\rm debris}$
in this time (or, equivalently, a fraction $f=[{m_{\rm debris}/ 
(m_{\rm debris}+m_{\rm sat}})]^{1/n}$ of its bound mass on each passage)
we can use equation (\ref{dndp}) to
calculate the matter distribution $dN/d\Psi$ along the
path of the satellite. This can be normalized
to give a mass surface density in $M_{\odot}/({\rm degree})^2$
\begin{equation}
	\Sigma(\Psi)={m_{\rm debris}\over w} {dN \over d\Psi}
=m_{\rm debris}{R_p \over r_{\rm tide}}  {dN \over d\Psi}
\end{equation}
where $w$ is the width of the debris trail (see eq. [\ref{width}]).
(For simplicity, we have adopted the viewpoint of an observer at
the Galactic center in this and all subsequent calculations.)

\begin{table}
\begin{center}
\begin{tabular}{cccccccc}
Satellite & mass        & L           	  & R   & $R_a/R_p$ & $T_{\Psi}$ & debris scale &
proper \\
          & ($M_{\odot}$) & ($L_{\odot}$) 	  & (kpc) & & (Gyrs) 	     & (width $\times$ 
length) & motion? \\
\hline
\hline
LMC       & $10^{10}$   & $2\times 10^{9}$  	& 31.9& 2.2 &1.5 & $17^{\circ}
\times 360^{\circ}$ & a 
\\
\hline
Sgr       & $10^8$      & $2\times 10^7$        & 12.7& 3.1 & 0.7 & $6^{\circ}
\times 360^{\circ}$ & b 
\\
\hline
Scul      & $1.4\times 10^7$ & $1.6\times 10^6$ & 52.0& 2.4 & 2.6 & $2^{\circ}\times 36^{\circ} $ & c\\
\hline
UMinor	  & $3.9\times 10^7$ & $3.0\times 10^5$ & 45.7& 2.0 & 2.0 & $3^{\circ}\times
90^{\circ} $ & d \\
\hline
Draco	  & $5.2\times 10^7$ & $2.5\times 10^5$ & 76.0& - &2.6 & $3^{\circ}\times
60^{\circ}$  & - \\
\hline	  
Carina	  & $1.1\times 10^7$ & $2.9\times 10^5$ & 89.0& - &3.0 & $2^{\circ}\times
30^{\circ} $ & - \\
\hline
Sextans	  & $2.6\times 10^7$ & $8.3\times 10^5$ & 91.0& - &3.1 & $2^{\circ}\times
45^{\circ} $ & -\\
\hline
Fornax	  & $1.2\times 10^8$ & $2.5\times 10^7$ & 133.0&- &4.5 & $3^{\circ} \times
36^{\circ} $ & - \\
\end{tabular}
\end{center}
\caption
{Parameters of Milky Way satellites. Columns:
(1) name; (2) mass; (3) luminosity; (4) 
distance from Galactic center;
(5) ratio of apocentric to pericentric distance;
(6) azimuthal time period; (7) approximate debris scale,
calculated from the width and length estimates given
by equations (\ref{width}) and (\ref{length}) ; (8) proper motion
references: a. \cite{jkl94}; b. \cite{i96}; c. \cite{s95}; d. \cite{s97}. 
Columns (4) and (5) give pericentric distances and
time periods integrated in the 
Galactic potential for those with proper motions.
Otherwise these give the current distance of the satellite 
and the time period for a circular orbit at that distance.
The mass, luminosity and current galactocentric distances are
from Jones, Klemola \& Lin (1994 --- LMC), 
Ibata, Irwin \& Gilmore (1994 --- Sgr) 
and Mateo et al. (1993 --- remaining dwarf spheroidals).
\label{satstab}
}
\end{table}

\begin{figure}
\begin{center}
\epsscale{1.0}
\plotone{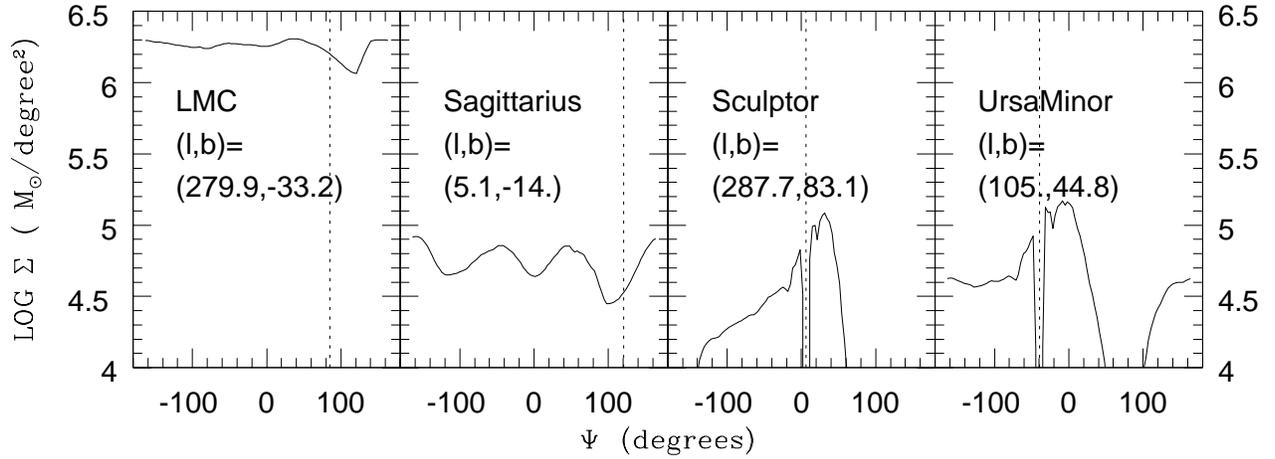}
\caption{Mass surface density distribution for debris
along the orbits of those
satellites with known proper motions, assuming each has lost 
half its mass in the last 10 Gyrs. 
The dotted lines show the current position of each satellite.
In the right hand panels,
the gap in the surface density at this point is the region that is
occupied by particles that are still
bound to the satellite
Note that $\Psi$ is measured from the Galactocentric, not heliocentric 
position.
\label{satsfig1}
}
\end{center}
\end{figure}

Figure \ref{satsfig1} shows the prediction for $\Sigma$
along the paths  (calculated from their proper-motions, see 
references in Table \ref{satstab})
of the LMC and the Sagittarius, Ursa Minor and Draco dwarf spheroidals
assuming $m_{\rm debris}=m_{\rm sat}$ for each satellite.
From estimates for their central
luminosity densities and mass-to-light ratios, (see, e.g.
\cite{ih95})
the central surface densities of these satellites are 
typically $> 10^8 M_{\odot}
/{\rm degree^2}$.
The current position of each satellite is given in
Galactic latitude and longitude $(l,b)$, and the corresponding value for 
the position $\Psi$ along the orbit is
indicated with the vertical dotted lines.
Note that since the extent of the debris' energy distribution
is only weakly dependent on the
instantaneous mass of the satellite ($\epsilon \propto m_{\rm sat}^{1/3}$)
the shape of these curves is nearly independent of $m_{\rm debris}$
so long as $m_{debris} \le m_{sat}$, and they 
can be simply scaled down
in amplitude for lower mass loss rates.

Debris from the LMC and Sagittarius
spreads over more than $2\pi$ in
$\Psi$ in 10 Gyrs, and in these cases 
the plot shows the result of summing over all mass which has the
same angular phase (i.e. assuming that the orbit is perfectly planar
and the same phase corresponds to the same line of sight).
Our prediction for the extent of the debris trail from the LMC
is consistent with the hypothesis that an ancient accretion
event is responsible for
the curious alignment of several 
dwarf spheroidal satellites (\cite{lb76}, 1982),
young globular clusters (\cite{lr92}; \cite{m94}; \cite{fp95}) and 
proper motions of some of these objects
(\cite{mpr96}) along the Great Circle define by the LMC
and the Magellanic Stream.
This calculation also confirms the idea that
a globular cluster that was previously a member of Sagittarius
(and several are currently associated with this dwarf
--- see \cite{da95}) could now be on the opposite
side of the Galaxy (\cite{fp95}) since the debris is expected to
encircle the sky within the lifetime of the Milky Way.

For any given luminosity function, we can 
use our estimate for $\Sigma$ to find the surface density of stars 
along the trail in apparent magnitude bins.
We illustrate this idea by considering horizontal branch (HB) stars,
which we expect to be useful tracers of debris trails.
Although not as bright as giant stars, they are easier to distinguish
from the foreground disk population using their blue colors.
Moreover, we can approximate their absolute magnitudes as
\begin{equation}
\label{hb}
	M_{\rm HB}=0.66+0.19[{\rm Fe/H}]
\end{equation}
(\cite{w92}), and therefore find their distances.
From equation (\ref{hb}), we know that the apparent magnitudes
of HB stars at distances of 25 -- 100 kpc lie in the range 17.5 -- 20.5.
In this example we will ignore any color information, and
simply contrast their number
density along each streamer
with star counts at the same Galactic latitude and
longitude $(l,b)$ and in the same apparent magnitude range, 
estimated from the Bahcall-Soneira Galactic model, $N_{\rm BS}$
(\cite{bs}).
We convert Figure \ref{satsfig1} 
into an estimate for the number of satellite
horizontal-branch stars per square degree via
\begin{equation}
	N_{\rm HB} =
	n_{\rm HB} {L_{\rm sat}\over m_{\rm sat}} \Sigma(\Psi),
\end{equation}
where $n_{\rm HB}$ is the number of horizontal branch stars per unit
luminosity for the satellite's population, and 
$L_{\rm sat}$ is the total luminosity 
of the satellite.
To estimate a typical value for $n_{\rm HB}$ we note that
the observed ratio of the number of HB stars to RGB
stars with magnitudes less than the horizontal branch lies in the range
$n_{\rm HB}/n_{\rm RGB}\sim1-2$ (\cite{mig90}; \cite{bcf85}). 
Then, using \cite{bv92} 
isochrones for an 8.0 Gyr, $[{\rm Fe/H}]=-1.03$ cluster as a representative
population, we find $n_{\rm RGB}=1/528$,
which implies $n_{\rm HB}\sim 0.002-0.004$.  

\begin{figure}
\begin{center}
\epsscale{1.0}
\plotone{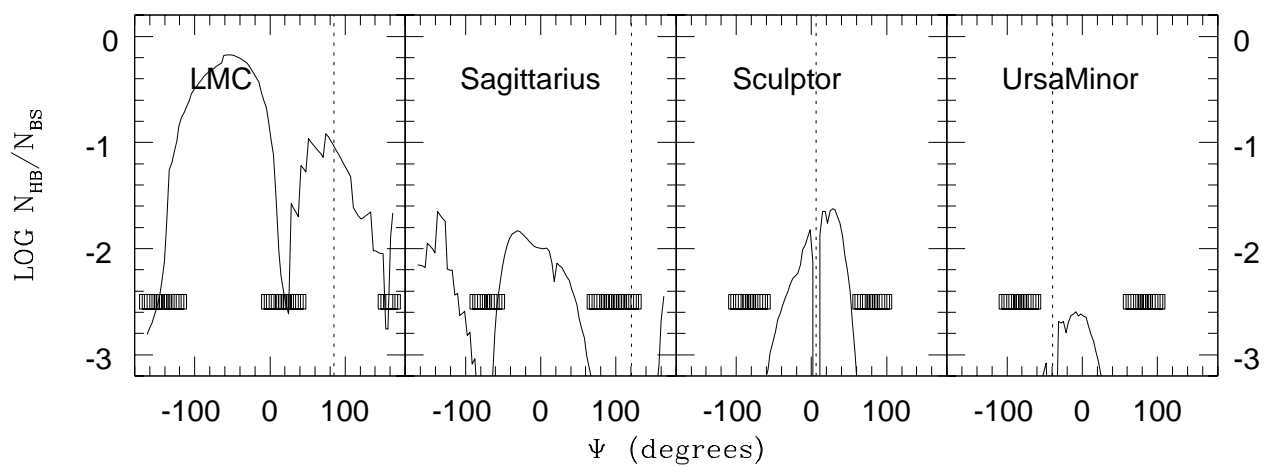}
\caption{
Conversion of Figure \ref{satsfig1} to a
comparison of the density of horizontal branch
stars, $N_{\rm HB}$,
with the background star counts from the Bahcall-Soneira
model for the Galaxy, $N_{\rm BS}$. 
The squares in these panels indicate where the
orbit is within 20 degrees of the Galactic disk and where the 
signature of the debris is most likely
to be swamped by foreground objects.
The dotted lines show the current position of each satellite.
In the right hand panels,
the gap in the surface density at this point is the region that is
occupied by particles that are still
bound to the satellite
\label{satsfig2}
}
\end{center}
\end{figure}

In Figure \ref{satsfig2} we plot
$N_{\rm HB}/N_{\rm BS}$ for $n_{\rm HB}=0.003$. 
The squares mark positions along each satellite's orbit that are less than
20 degrees from the Galactic plane, where the background counts
are largest.
The comparison suggests that even for large mass loss rates, only the tidal
streamers from the LMC would be 
detectable as local over-densities in
star counts, though the debris from the other satellites might be found
by integrating the star counts along the suspected orbital path
(as proposed in JHB) or
using additional color, kinematic or distance information.
These examples also imply that tidal
streamers from such minor accretion events in external galaxies
are likely to be unobservable.

\subsection{How Long Will We See a Satellite After Disruption?}

\begin{figure}
\begin{center}
\epsscale{0.5}
\plotone{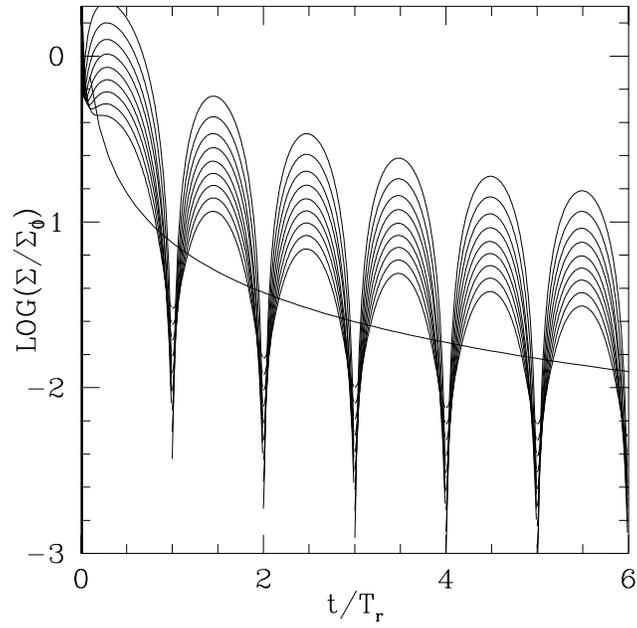}
\caption{Evolution of the central surface density of a remnant
following the complete destruction of a satellite, for orbits with the
same energy and eccentricities as in Figure \ref{densfig}
Pericenter corresponds  to the dip in each curve.
\label{sigfig}
}
\end{center}
\end{figure}

The central surface density of a remnant at time $t$ after 
the destruction of a satellite is given by
\begin{equation}
\label{sig}
        {\Sigma \over \Sigma_0}= m_{\rm sat} {dN \over d\Psi}{R_p\over r_{\rm tide}} 
{1 \over \Sigma_0}
={1 \over 2 d} {1 \over t\dot{\Psi}(t)},
\end{equation}
where $\Sigma_0= m_{\rm sat} R_p^2/r_{\rm tide}^2$ is the approximate surface
density of the satellite as it disrupts.
This is plotted in Figure \ref{sigfig} 
for a circular orbit (bold) and orbits with eccentricities
spanning the range $R_p/R_a=0.07-1.0$, 
where $R_a$ is the apocentric distance (as in Figure \ref{densfig}).
In all cases, $\Sigma/\Sigma_0$ oscillates between apocenter and pericenter,
but with a characteristic drop in amplitude by more than a 
factor of 10 within the first two radial periods ($\sim \sqrt{2}T_{\Psi}$)
--- in agreement with the results from Kroupa's (1997)
simulations. 
For objects that have already been destroyed, 
we may take the surface densities of the dwarf spheroidals observed today
(which are 10--100 times the background star counts --- see \cite{ih95})
as upper limit to $\Sigma_0$.
This implies that
the signatures of such events as a local
overdensity in star counts would be clear only for the time period of
the orbit (i.e. $\sim 1$ Gyr, see eq. [\ref{tee}]).
However, the amplitude of the oscillation only decays linearly with
time. This suggests that either a matched filtering technique, where star
counts are integrated over some region (e.g. \cite{jhb96}),
or a search including kinematic information
could effectively find debris for the lifetime of the
Galaxy (tens of orbital periods).

\subsection{How Lumpy is the Halo?}

\begin{figure}
\begin{center}
\epsscale{0.75}
\plotone{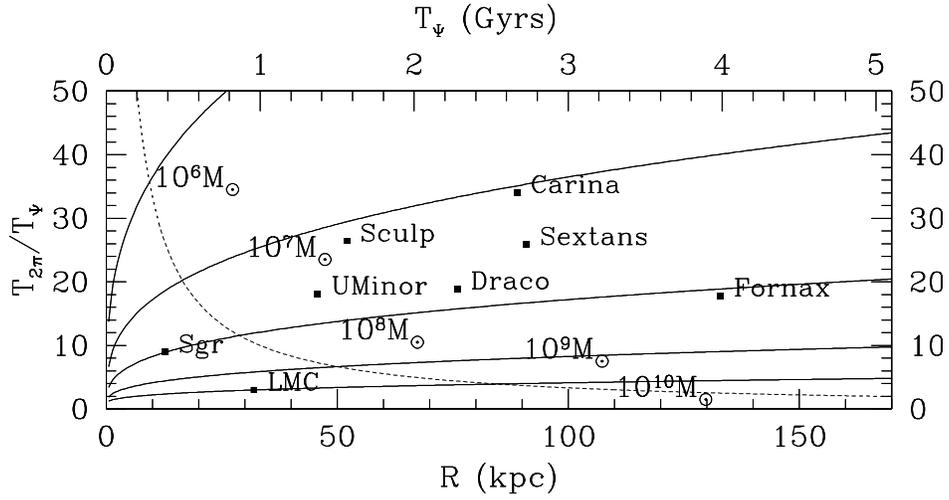}
\caption{
The time taken for debris to
disperse over a full Great Circle (in units of the
azimuthal time period of the orbit), as a function of 
distance and for a variety of satellite masses (solid lines).
The dotted line is the curve $T_{2\pi}=10$ Gyears.
The Galactic satellites are plotted either at their current distance or
their calculated pericentric distance (for those with proper motions).
The equivalent time period for a circular orbit at each distance is
shown along the top axis.
For example, this plot implies that 
it will take $(T_{2\pi}/T_{\Psi} \sim 19) \times (T_{\Psi} \sim
2.3 \> {\rm Gyrs}) 
= 43.7 \> {\rm Gyrs}$
for the tidal streamers from Draco to encircle the Galaxy.
\label{tdispfig}
}
\end{center}
\end{figure}

We can
estimate a timescale to characterize the rate of debris dispersal by
calculating how long it will take the streamers
from a satellite to extend $2\pi$ along its orbit:
\begin{equation}
        {T_{2\pi}\over T_{\Psi}}={\tau(2\epsilon) \over \tau(2\epsilon)-1}
                                ={1 \over 1 - \exp (-\epsilon/v_h^2)}
		={1 \over 1 - \exp(-(2Gm/R_p v_{\rm circ}^2)^{1/3})}.
\end{equation}
Figure \ref{tdispfig} 
plots $T_{2\pi}/T_{\Psi}$ as a function of pericentric radius 
for various satellite masses, and for each of the Milky Way's satellites
(at either their pericentric or current positions for those with or
without proper motion data respectively).
The upper axis shows $T_{\Psi}$ for a circular orbit at
radius $R_p$.
The plot confirms the result of \S3.1
that debris from the LMC and Sagittarius is expected to spread more than
$2\pi$ along their orbits over the lifetime of the Galaxy.

\begin{figure}
\begin{center}
\epsscale{0.75}
\plotone{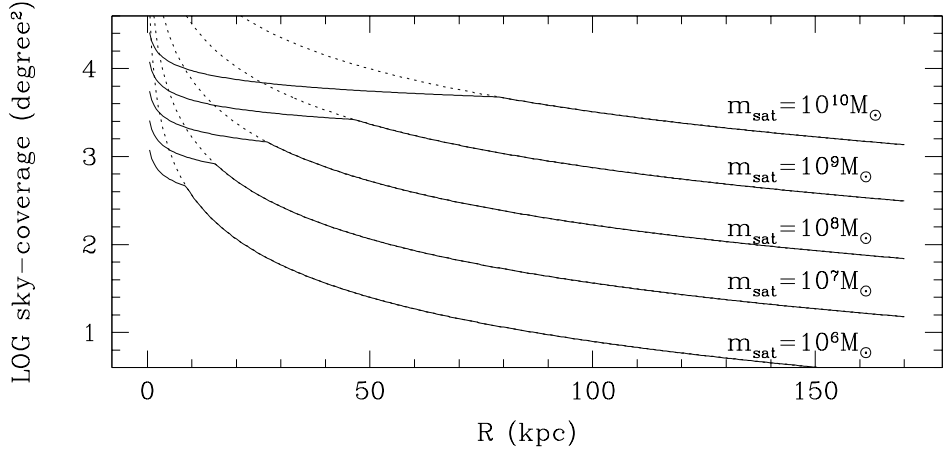}
\caption{Each curve shows
the fractional sky coverage of debris from a satellite
of mass $m_{\rm sat}$ after 10 Gyrs, as a function of its
Galactocentric distance. The dotted lines show the behavior if
the streamer is not assumed to overlap itself when it
extends more than $2\pi$ along the satellite's orbit.
\label{fracfig1}
}
\end{center}
\end{figure}

We can use this timescale to find the approximate angular length $L$
of a streamer
from a satellite that has lost mass during the last 10 Gyrs:
\begin{equation}
\label{length}
	L={10 {\rm Gyr} \over T_{\Psi}} { 2\pi \over T_{2\pi}/T_{\Psi}},
\end{equation}
and this is included in Table \ref{satstab}.
Combining this with the width of the streamer (eq. [\ref{width}])
and approximating $T_{\Psi}$ with
the time period of a circular orbit at the pericentric position, we
can estimate the sky coverage $w\times L$
of debris from satellites of various masses and pericentric distances, 
as shown in Figure \ref{fracfig1}.

The lumpiness of halos with different accretion histories
can be assessed by superimposing an ensemble of these trails
and calculating how much of the sky they cover. 
We assume that $M_{\rm acc}$ of the mass of the halo
has been accreted in the form of $M_{\rm acc}/m_{\rm sat}$ satellites of
mass $m_{\rm sat}$, to form a density distribution that follows that
of the stellar halo (where $\rho \propto r^{-3.5}$, see \cite{f96}).
Then we integrate the sky coverage of each individual
trail over the number density of accreted satellites to find the
expected number of trails $n_{\rm trails}$ 
intersecting any random line of sight
(or, equivalently, how much of the sky is covered).

\begin{figure}
\begin{center}
\epsscale{0.75}
\plotone{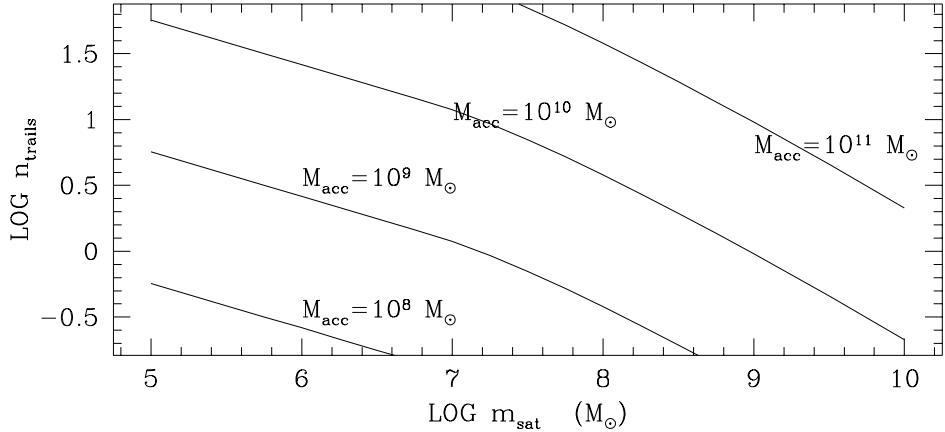}
\caption{
The curves 
show the number of trails
observed along a random tine of sight (or, equivalently, the
fraction of the sky covered)
for total accreted mass $M_{\rm acc}$ as a function of
the mass of the accreted satellites.
\label{fracfig2}
}
\end{center}
\end{figure}

Figure \ref{fracfig2} shows $n_{\rm trails}$
as a function of satellite mass for $M_{\rm acc}=10^8,
10^9, 10^{10}, 10^{11} M_{\odot}$
(see also \cite{t93} for a simple --- and elegant --- version of this
calculation).
Of course, we know neither the accretion history of the Galaxy
nor the primordial spectrum of satellite masses.
However, semi-analytic calculations suggest that 
many of the 
Galactic globular clusters will disrupt
in the next Hubble time and that those we see today
are the remnants of a much larger primordial population
(e.g. \cite{mw96}; \cite{go97}). 
From \S 3.2, we would not expect to see trails from
such disruption events for more than a few orbital periods
in star counts alone. 
However,
if we suppose that $\sim 100$ similar
objects (with masses of $10^5$ -- $10^6 M_{\odot}$) 
have been accreted in the halo in the last Hubble time, then
Figure \ref{fracfig2} predicts that their debris trails would
cover 10\% of the sky.
This significant sky coverage encourages the idea that many could be
detected with kinematic surveys of halo stars along random
lines of sight.

\section{Microlensing in a Lumpy Halo}

In a recent paper, \cite{z97} proposed several different 
scenarios in which the superposition of a
debris trail between us and the Magellanic Clouds 
could explain all the microlensing events observed by the MACHO team
(\cite{aea96}).
In a field 2$^{\circ}$ northwest of the LMC bar,
\cite{zl97} find tantalizing evidence for such a foreground population
of a suitable density and distance: 5\% of the clump
giants in their survery are 0.9 magnitudes brighter than the mean
(note that \cite{g97} questions the conclusion that this population could
indeed account for the microlensing events).
With the tools developed in this paper we can
re-examine the likelihood of each of Zhao's scenarios,
and hence the interpretation of Zaritsky \& Lin's clump stars
as tidal debris.

In Zhao's first scenario, he assumes that the Magellanic plane 
(defined by the LMC and the Magellanic stream) 
is populated by a stream of debris, with lensing mass $10^{10} M_{\odot}$,
uniformly distributed along a great circle ---
equivalent to the model we
discussed in \S 3.1 for tidal streamers from the LMC. 
Our formalism predicts that the streamers
would cover an area $17\times 360$ square-degrees (see Table
\ref{satstab}). 
If the debris has the same mass-to-light ratio as the LMC, this would
correspond to the luminosity of the entire stellar halo ($\sim 10^9 L_{\odot}$,
see \cite{f96}) in only 10\% of the sky and would produce an
over-density 
in star-counts across the Magellanic Plane at high Galactic latitude
of 10-100\% (see Figure \ref{satsfig2}).
It seems unlikely that such a significant 
population could have been missed in previous studies
since the number-counts predicted by the
Bahcall-Soneira model have been found to provide a reasonable fit
to observations in all fields to which they have been
applied (\cite{b86}; \cite{rm93}).

As an alternative, Zhao suggests debris just covering the
extent of the clouds, either $10^{10}M_{\odot}$
immediately in front of the clouds ($\sim 45$ kpc from the Galactic center), 
or $10^8-10^9M_{\odot}$
at the Galactocentric distance of Sagittarius (16 kpc).
In both cases, the timescale for such debris to disperse over an orbit is
$\sim 3$Gyrs (from the positions of the LMC and Sagittarius in
Figure \ref{tdispfig}) which suggests that 
such an alignment over just $10^{\circ}$ would only last for $< 10^8$years.
Thus, this requires the destruction of an object of either mass within
the last
$10^8$ years, and the alignment of its debris in just 1/400th of the sky. 
We conclude that, though not impossible, 
these scenarios are also unlikely.

\section{Summary and Discussion.}

We have outlined a method that
can construct debris trails,
using negligible computational resources, for any specified
satellite mass and orbit and in most parent
potentials, requiring only
the assumption of a timescale over which the satellite is destroyed
(see \S 2.2).

The debris distributions generated by our approach 
demonstrate a remarkable agreement 
with those calculated from the positions of unbound particles
at the end of simulations run in a fixed Galactic potential
using equivalent satellite and orbit parameters
(\S 2.2.3 and Figure \ref{npsifig}).
This success encourages its use as a tool for
understanding the evolution of tidal streamers from
the accretion of small ($m_{\rm sat}/M_{\rm MW} < 1/100$) satellites 
after the potential fluctuations due to 
the initial formation of the Galaxy have died down, and for which
the limitations mentioned in the
following paragraph do not apply.

The approximations inherent in this prescription (see \S 2.3)
imply that it is not 
well-suited to generating debris distributions if the mass loss rate
is nearly constant (e.g. if the satellite is on a near circular orbit) or
if the satellite's orbit is evolving due to dynamical friction with the
parent galaxy.
Both of these effects could conceivably be taken into
account with a more sophisticated implementation of our technique.
The method is clearly inapplicable to problems where
the parent's potential is either triaxial or fluctuating as the
result of a recent merger, such that the debris might be expected to diverge
further from the satellite's orbit.

\acknowledgements 
I thank Roeland van der Marel and John Bahcall for invaluable comments on 
the manuscript, 
Lars Hernquist and Mike Bolte for the collaboration 
that provided the inspiration (and simulations) for this project,
and Steve Majewski for many useful discussions.
The simulations used 
here were performed at the Pittsburgh Supercomputing Center.
This work was supported by the Monell Foundation and the Institute for Advanced
Study SNS Membership Fund.

\appendix
\section{Simulations}

The following is a brief sketch of the method and parameter ranges
for the simulations, which are described fully in
\cite{jhb96}.

In each simulation we represent the Milky Way  by a rigid,
three-component potential. 
The disk is described by a Miyamoto-Nagai potential (\cite{mn75}), 
the spheroid by a Hernquist (1990)
potential and the halo by a logarithmic potential:
\begin{eqnarray}
\label{mw}
        \Phi_{\rm disk}&=&-{GM_{\rm disk} \over
                 \sqrt{R^{2}+(a+\sqrt{z^{2}+b^{2}})^{2}}}, \cr
        \Phi_{\rm spher}&=&-{GM_{\rm spher} \over r+c}, \cr
        \Phi_{\rm halo}&=&v_{\rm halo}^2 \ln (r^{2}+d^{2}).
\end{eqnarray}
We take $M_{\rm disk}=1.0 \times 10^{11}, M_{\rm spher}=3.4 \times 10^{10}, 
v_{\rm halo}= 128, a=6.5, b=0.26, c=0.7$, and
$d=12.0$, where masses are in $M_{\odot}$, velocities are in km/s
and lengths are in  kpc. 

Each satellite is modeled with a collection of $10^4$
self-gravitating particles, initially
represented by a Plummer (\cite{p11}) model
\begin{equation}
        \Phi=-{Gm \over \sqrt{r^2+r_0^2}},
\end{equation}
where $m$ is the mass of the satellite and $r_0$ is its scale length.
During the simulations, the
mutual interactions of the satellite particles are calculated
using a self-consistent field code (\cite{ho92}),
with the Milky Way's influence calculated from the 
equation (\ref{mw}).

In this paper, we consider the state of the satellite
after 10 Gyrs of evolution in
twelve different simulations, whose  parameters are
outlined in Table \ref{modstab}.
The first four models (1--4) have satellite and orbit
parameters chosen at random to explore a range of possible outcomes. 
The remaining eight models 
all had the same orbit as Model 2, but with satellite parameters
chosen to change the rate of mass loss and debris dispersal.
In Models 5--8  the satellites 
have the same central density as Model 2, 
but different velocity dispersions.
In Models 9--12 the satellites 
have the same velocity dispersion as Model 2, 
but different central densities.

\newpage

\newpage

\begin{thebibliography}{DUM}

\bibitem[Alard 1996]{a96}
Alard, C. 1996, \apjlett, 458, L17

\bibitem[Alcock et al. 1996]{aea96}
Alcock, C. et al. 1996, astro-ph/9606165

\bibitem[Arnold \& Gilmore 1992]{ag92} 
Arnold, R. \& Gilmore, G. 1992, \mnras, 257, 225

\bibitem[Bahcall 1986]{b86}
Bahcall, J. N. 1990, \araa, 24, 577

\bibitem[Bahcall \& Soneira 1980]{bs}
Bahcall, J. N. \& Soneira, R. M. 1980, \apjs, 44, 73

\bibitem[Bergbusch \& VandenBerg's (1992)]{bv92}
Bergbusch, P. A. \& VandenBerg, D. A. 1992, \apjs, 81, 163

\bibitem[Binney \& Tremaine 1987]{bt}
 Binney, J. \& Tremaine, S. 1987, Galactic Dynamics
(Princeton University Press, Princeton)

\bibitem[Buonanno Corsi \& Fusi Pecci 1985]{bcf85}
Buonanno, R., Corsi, C. E. \& Fusi Pecci, F. 1985, \aap, 145, 97

\bibitem[Da Costa \& Armandroff 1995]{da95}
Da Costa, G. S. \& Armandroff, T. E. 1995, \aj, 109, 2533

%\bibitem[Dubinski \& Carlberg, 1991]{dc91}
%Dubinski, J. \& Carlberg, R. G. 1991, \apj, 378, 496

\bibitem[Eggen 1965]{e65}
Eggen, O.J. 1965, in Galactic Structure, eds. A. Blaauw \& M. Schmidt
(Chicago, University of Chicago Press), p. 111

%\bibitem[Eggen, Lynden-Bell \& Sandage (1962)]{els} 
%Eggen, O. J. Lynden-Bell, D.\& Sandage, A. R. 
%1962, \apj, 136, 748

\bibitem[Evans \& Kochanek 1989]{ek89}
Evans, C. R.\& Kochanek, C. S. 1989, \apjlett, 346, L13

\bibitem[Fahlman et al. 1996]{fmrts96}
Fahlman, G. G., Mandushev, G., Richer,H. B., Thompson, I. B. \&
Sivaramakrishnan, A. 1996, \apjlett, 459, L65

\bibitem[Freeman 1996]{f96}
Freeman, K. C. 1996, in The Formation of the 
Galactic Halo: Inside and Out, ASP Conf. Ser. Vol 92, eds. H. Morrison \&
A. Sarajedini, (ASP, San Francisco), p. 3 

\bibitem[Fusi-Pecci et al. 1995]{fp95}
Fusi Pecci, F., Bellazzini, M., Cacciari, C. \& Ferraro, F. R. 1995,
\aj, 110, 1664

\bibitem[Gnedin \& Ostriker 1997]{go97}
Gnedin, O. \& Ostriker, J. P. 1997, \apj, 474, 223

\bibitem[Gould 1997]{g97}
Gould, A. 1997, astro-ph/9709263

\bibitem[Gould et al. 1992]{ggrf92}
Gould, A., Guhathakurta, P., Richstone, D. \& Flynn, C. 1992, \apj, 
338, 345

\bibitem[Grillmair et al. 1995]{g95} 
Grillmair, C. J., Freeman, K. C., Irwin, M. \& Quinn, P. J. 1995, \aj,
109, 2553

\bibitem[Hernquist 1990]{h90} 
Hernquist, L. 1990, \apj, 356, 359

\bibitem[Hernquist \& Ostriker 1992]{ho92} 
Hernquist, L. \& Ostriker, J.  P. 1992, \apj, 386, 375

%\bibitem[Hesser, et al. 1996]{h96}
%Hesser, J. E. et al., 1996 J. Korean Astro. Soc., 
%29, 1, astro-ph/9703048

\bibitem[Ibata, Gilmore \& Irwin 1994]{igi94}
Ibata, R. A., Gilmore, G. \& Irwin, M. J. 1994, \nat, 370, 194

\bibitem[Ibata et al. 1996]{iwgis96} 
Ibata, R. A., Wyse, R. F. G., Gilmore, G., Irwin, M. J. \& Suntzeff, N. B.
1996, astro-ph/9612025, to appear in \aj

\bibitem[Irwin et al. 1990]{i90}
Irwin, M. J., Bunclark, P. S., Bridgeland, M. T. \& McMahon, R. G. 1990,
\mnras, 244, 16P

\bibitem[Irwin \& Hatzidimitriou 1995]{ih95}
Irwin, M. J. \& Hatzidimitriou, D. 1995, \mnras, 277, 1354

\bibitem[Irwin et al. (1996)]{i96}
Irwin, M. , Ibata, R., Gilmore, G., Wyse, R. \& Suntzeff, N.
1996, in The Formation of the 
Galactic Halo: Inside and Out, ASP Conf. Ser. Vol 92, eds. H. Morrison \&
A. Sarajedini, (ASP, San Francisco), p.84 

\bibitem[JHB]{jhb96} 
Johnston, K. V., Hernquist, L. \& Bolte, M. 1996, \apj, 465, 278 (JHB)

\bibitem[Johnston \& Majewski 1997]{jm97}
Johnston, K. V.  \& Majewski, S. R. 1997,  in preparation

\bibitem[Johnston, Spergel  \& Hernquist 1995]{jsh95} 
Johnston, K. V., Spergel, D. N.  \& Hernquist, L. 1995, \apj, 451, 598

\bibitem[Jones, Klemola  \& Lin (1994)]{jkl94}
Jones, B. F., Klemola, A. R.  \& Lin, D. N. C. 1994, \aj, 107, 1333

%\bibitem[Katz 1991]{k91}
%Katz, N. 1991, \apj, 368, 325

\bibitem[King (1962)]{k62} 
King, I. R. 1962, \aj, 67, 471

\bibitem[Klenya et al. (1997)]{kea97}
Klenya, J., Geller, M., Kenyon, S.  \& Kurtz, M. 1997, \aj, 113, 624

\bibitem[Kochanek 1994]{k94}
Kochanek, C. S. 1994, \apj, 422, 508

\bibitem[Kroupa 1997]{k97}
Kroupa, P. 1997, New Astronomy,  in press

%\bibitem[Kuhn 1993]{k93}
%Kuhn, J. R. 1993, \apjlett, 409, L13

\bibitem[Kuhn, Smith  \& Hawley 1996]{ksh96}
Kuhn, J. R., Smith, H. A.  \& Hawley, S. L. 1996, \apjlett, 469, L93

\bibitem[Kunkel \& Demers (1977)]{kd77}
Kunkel, W.E. \& Demers, S. 1977, \apj, 214, 21

\bibitem[Lin  \& Richer 1992]{lr92}
Lin, D. N. C.  \& Richer, H. B. 1992, \apj, 388, L57

\bibitem[Lynden-Bell 1976]{lb76} 
Lynden-Bell, D. 1976, \mnras, 174, 695

\bibitem[Lynden-Bell 1982]{lb82} Lynden-Bell, D. 1982, Observatory, 102, 202

\bibitem[Lynden-Bell  \& Lynden-Bell 1995]{lb295}
Lynden-Bell, D.  \& Lynden-Bell, R. M. 1995, \mnras, 275, 429

\bibitem[McGlynn 1990]{m90} 
McGlynn, T. A. 1990, \apj, 348, 515

\bibitem[Maddox et al. 1990]{apm}
Maddox, S. J., Sutherland, W. J., Efstathiou, G.  \& Loveday,
J. 1990, \mnras, 243, 692

%\bibitem[Majewski 1993]{m93} 
%Majewski, S. R. 1993, \araa, 31, 575

\bibitem[Majewski 1994]{m94} 
Majewski, S. R. 1994, \apjlett, 431, L17

\bibitem[Majewski, Hawley \& Munn 1996]{mhm96}
Majewski, S. R., Hawley, S. L. \& Munn, J. A. 1996, in The Formation of the 
Galactic Halo: Inside and Out, ASP Conf. Ser. Vol 92, eds. H. Morrison  \&
A. Sarajedini, (ASP, San Francisco), p. 119

\bibitem[Majewski, Munn \& Hawley 1994]{mmh94}
Majewski, S. R., Munn, J. A.  \& Hawley, S. L. 1994, \apj, 427, L37

\bibitem[Majewski, Munn \& Hawley 1996]{mmh96}
Majewski, S. R., Munn, J. A. \& Hawley, S. L. 1996, \apj, 459, L73

\bibitem[Majewski, Phelps  \& Rich 1996]{mpr96} 
Majewski, S. R., Phelps, R. L.  \& Rich, R. M. 1996, to
appear in The History of the 
Milky Way and its Satellite Systems, ASP Conf. Ser, eds.
A. Burkert, D. H. Hartmann  \& S. R. Majewski.

\bibitem[Mateo et al. 1996]{m96}
Mateo, M., Mirabel, N., Udalski, A., Szyma\'{n}ski, M., Kubiak, M.,
Krzemi\'{n}ski, W.  \& Stanek, K. Z. 1996, \apjlett, 458, L13

\bibitem[Mateo et al. 1996]{mea93}
Mateo, M., Olszewski, E.W., Pryor, C., Welch, D.L. \& Fischer, P. 1993, 
\aj, 105, 510

\bibitem[Metcalf  \& Silk (1996)]{ms96}
Metcalf, B.  \& Silk, J. 1996, \apj, 464, 218

\bibitem[Mighell 1990]{mig90}
Mighell, K. 1990, \aaps, 82, 207

\bibitem[Miyamoto  \& Nagai 1975]{mn75} 
Miyamoto, M.  \& Nagai, R. 1975, \pasj, 27, 533

%\bibitem[Monet 1995]{usno}
%Monet, D. 1995, U.S. Naval Observatory compilation.

\bibitem[Moore  \& Davis 1994]{md94}
Moore, B. \& Davis, M. 1994, \mnras, 270, 209

\bibitem[Murali \& Weinberg 1996]{mw96}
Murali, C. \& Weinberg, M. D. 1996, astro-ph/9610229

%\bibitem[Navarro, Frenk \& White 1996]{nfw}
%Navarro, J. F., Frenk, C. S. \& White, S. D. M.  1996, \apj 462, 563

%\bibitem[Norris 1994]{n94}
%Norris, J. 1994, \apj, 431, 645

%\bibitem[Norris 1996]{n96}
%Norris, J. 1996, in The Formation of the 
%Galactic Halo:Inside and Out, ASP Conf. Ser. Vol 92, eds. H. Morrison \&
%A. Sarajedini, (ASP, San Francisco), p. 14

\bibitem[Oh, Lin \& Aarseth 1995]{ola95}
Oh, K.S., Lin, D.N.C. \& Aarseth, S.J. 1995 \apj, 442, 142

%\bibitem[Paczy\'{n}ski (1986)]{p86}
%Paczy\'{n}ski, B. 1986, \apj, 304, 1

\bibitem[Pennington et al. 1993]{poss1}
Pennington, R. C., Humphreys, R. M., Odewahn, S. C., Zumach, W.
\& Thurmes, P. M. 1993, \pasp, 105,521

\bibitem[Piatek \& Pryor 1995]{pp95} 
Piatek, S. \& Pryor, C. 1995, \aj, 109, 1071

\bibitem[1911]{p11} 
Plummer, H. C. 1911, \mnras, 71, 460

\bibitem[Reid \& Majewski 1993]{rm93}
Reid, N. \& Majewski, S. R. 1993, \apj, 409, 635

\bibitem[Roche 1847]{r47}
Roche, E. A. 1847, Academie des Sciences et Lettres de Montpellier,
Memoirs de la Section des Sciences, Vol. 1, 243-262

\bibitem[Sackett 1997]{penny}
Sackett, P. D. 1997, \apj, 483, 103

%\bibitem[Sandage 1990]{s90}
%Sandage, A. 1990, \jrasc, 84, 70

\bibitem[Schweitzer, Cudworth, Majewski \& Suntzeff (1995)]{s95}
Schweitzer, A.  E., Cudworth, K.  M., Majewski, S. R. \& Suntzeff, N. B.
1995, \aj, 110, 2747

\bibitem[Schweitzer, Cudworth \& Majewski (1997)]{s97}
Schweitzer, A.  E., Cudworth, K.  M. \&  Majewski, S. R. 
1997, \aj, submitted

\bibitem[Scholz \& Irwin (1993)]{si93}
Scholz, R. D. \& Irwin, M. J. 1993, I.A.U Symposium 161, p.535

%\bibitem[Searle \& Zinn (1978)]{sz} Searle, L. \&  Zinn, R.  
%1978, \apj, 225, 357

\bibitem[Sommer-Larsen \& Christiansen 1987]{slc87} 
Sommer-Larsen, J. \& Christiansen, P. R. 1987, \mnras, 225, 499

\bibitem[Sridhar \& Tremaine 1992]{st92}
Sridhar, S. \& Tremaine, S. 1992, Icarus, 95, 86

\bibitem[Tremaine 1993]{t93}
Tremaine, S. 1993 in Back to the Galaxy eds. S. S. Holt \& F. Verter,
(AIP Conf. Proc. : New York), p.599

\bibitem[Unavane, Wyse \& Gilmore (1996)]{uwg96}
Unavane, M., Wyse, R.F.G. \& Gilmore, G. 1996 \mnras, 278, 727

\bibitem[Velazquez \& White 1995]{vw95}
Velazquez, H. \& White, S.D. 1995 \mnras, 275, L23

\bibitem[Walker 1992]{w92}
Walker, A. R. 1992, \apjlett, 390, L81

\bibitem[Weir, Fayyad \& Djorgovski 1995]{poss2}
Weir, N., Fayyad, U. M. \& Djorgovski, S. G. 1995 \aj, 109, 2401

%\bibitem[White 1996]{w96}
%White, S. D. M. 1996, astro-ph/9602054

\bibitem[Zaritsky \& Lin (1997)]{zl97}
Zaritsky, D. \& Lin, D.N.C. 1997, \apj, in press

\bibitem[Zhao (1997)]{z97}
Zhao, H. S. 1997, astro-ph/9703097

\end{thebibliography}
\end{document}